\documentclass{aastex631}

\usepackage{xcolor}

\usepackage{ifthen}
\newcommand{\alphacen}[1]{%
    \ifthenelse{\equal{#1}{short}}{$\alpha~Cen$}{%
    \ifthenelse{\equal{#1}{full}}{$\alpha~$Centauri}{%
    \textcolor{red!55!red}{Error: Invalid argument}}%
}}

\usepackage{enumitem}

\received{December 17, 2024}
\revised{Jan 24, 2025}
\accepted{January 31, 2025}

\begin{document}

\title{A Case Study of Interstellar Material Delivery: \alphacen{full}}

\author[0000-0001-8927-7708]{Cole R. Gregg}
\affiliation{Department of Physics and Astronomy \\
The University of Western Ontario \\
London, Canada}
\affiliation{Institute for Earth and Space Exploration (IESX) \\
The University of Western Ontario \\
London, Canada}

\author[0000-0002-1914-5352]{Paul A. Wiegert}
\affiliation{Department of Physics and Astronomy \\
The University of Western Ontario \\
London, Canada}
\affiliation{Institute for Earth and Space Exploration (IESX) \\
The University of Western Ontario \\
London, Canada}

\begin{abstract}
Interstellar material has been discovered in our Solar System, yet its origins and details of its transport are unknown.
Here we present \alphacen{full} as a case study of the delivery of interstellar material to our Solar System. \alphacen{full} is a mature triple star system that likely harbours planets, and is moving towards us with the point of closest approach approximately 28,000 years in the future.
Assuming a current ejection model for the system, we find that such material can reach our Solar System and may currently be present here. The material that does reach us is mostly a product of low ($<2$ km s$^{-1}$) ejection velocities, and the rate at which it enters our Solar System is expected to peak around the time of \alphacen{full}'s closest approach. If \alphacen{full} ejects material at a rate comparable to our own Solar System, we estimate the current number of \alphacen{full} particles larger than 100~m in diameter within our Oort Cloud to be $10^{6}$, and during \alphacen{full}'s closest approach, this will increase by an order of magnitude. However, the observable fraction of such objects remains low as there is only a probability of $10^{-6}$ that one of them is within 10~au of the Sun. A small number ($\sim 10$) meteors $> 100~\mu$m from \alphacen{full} may currently be entering Earth's atmosphere every year: this number is very sensitive to the assumed ejected mass distribution, but the flux is expected to increase as \alphacen{full} approaches.

\end{abstract}


\section{Introduction}\label{sec:Intro}



To date, there have been two discoveries of macroscopic objects from outside our Solar System: 1I/'Oumuamua \citep{Oumuamua_discovery} and 2I/Borisov \citep{Borisov_charac}. At smaller sizes, in situ dust detectors on spacecraft \citep{Grun1997,baguhl1995,altobelli2003} have also unambiguously detected interstellar particles. At intermediate sizes, the claims of interstellar meteors detected remain controversial. This is because usually the only indicator of the interstellar nature of a particle is its hyperbolic excess velocity, which is very sensitive to measurement error \citep{hajdukova_2020, hajdukova_2024}. 

In any case, the details of the travel of interstellar material as well as its original sources remain unknown. Understanding the transfer of interstellar material carries significant implications as such material could seed the formation of planets in newly forming planetary systems \citep{griperavn19,Moro-Mart_2022}, while serving as a medium for the exchange of chemical elements, organic molecules, and potentially life's precursors between star systems - panspermia \citep{griperavn19, Adams_Napier_2022, osmanov_2024, smith_2024}.

Here we aim to increase our understanding of interstellar transport by performing a case study of one particular nearby star system, \alphacen{full} (\alphacen{short}), and focusing on transfer within the near term (last 100~Myr). The fundamental questions guiding this study are: Can \alphacen{short} plausibly be ejecting material at the current time, and if so, would we expect this material to arrive at our Solar System? What would be the expected characteristics of this material, including arrival direction, velocity, and flux?

The first question is whether \alphacen{short} can reasonably be expected to be ejecting material at the current time. As the system is mature, we would expect much of its original protoplanetary disk to have dissipated, with some of the mass possibly retained in asteroid/Kuiper belts, or an Oort cloud (OC). Our simulations examine the relatively recent past (astronomically speaking, $\sim100~$Myr), well after any planet-forming phase and the disassociation of the birth cluster \alphacen{short} formed in (typical lifetimes are $\lesssim 100s~$Myr \citep{adams_birth_2010}). Therefore, the primary ejection mechanisms of interest are gravitational scattering of leftover planetesimals by the stars and/or planets within the system, as well as the loss of distant OC members to galactic tidal stripping. Work much along these lines was performed by \cite{Zwart2021}, who modeled the evolution of OCs around the 200 nearest \textit{Gaia} stars for $1~$Gyr in the past to determine the density of interstellar objects (ISO) around the Sun, though \alphacen{short} is not discussed in detail in that work. Here we seek to extract the specific details of the material delivery from that system near the present time, though we do not model the process of ejection in detail, but rather adopt a suitable ejection velocity distribution for the system.

Assuming that \alphacen{short} is currently ejecting material, we find that such material can reach our Solar System and may currently be present here. This system is a good choice for this kind of case study for several reasons:

\begin{itemize}
    \item It is the closest star to our Solar System, at $1.34~$pc \citep{akeson_2021}. Its proximity increases the likelihood that material from this system can reach us.
    \item It is approaching our Solar System at $22~{\rm km~s}^{-1}$ \citep{RV_simbad_evans1967revision, kervella_proxima_2017, SIMBAD} and will pass within $200,000~$au of the Sun in 28,000 years. Thus we can expect that the amount of material delivered to us is increasing as the effective cross-section of the Solar System increases (see Section \ref{sec:Results} for more details).
    \item This is a mature (5 Gyr age \citep{akeson_2021, joyce_2018, thevenin_2002}) triple star system that likely harbours planets. Though mature star systems likely eject less material than those in their planet-forming years, the presence of multiple stars and planets increases the likelihood of gravitational scattering of members from any remnant planetesimal reservoirs, much as asteroid or comets are currently being ejected from our Solar System.
    \item Two of the stars, \alphacen{short} $A$ and $B$, are Sun-like stars (see below for details). Their larger than typical stellar mass suggests that they likely formed from a more massive than typical protoplanetary disk, which might allow more mass to remain in unaccumulated form. In particular, the system might have developed an OC, which results from gravitational scattering of planetesimals from Neptune mass planets \citep{saf72,duncan1987, tre93} which would provide a source of macroscopic bodies to eject via mechanisms much like those seen in our Solar System today (see Section \ref{sec:MassDensity}).
    
\end{itemize}

Though we will not model the ejection process from \alphacen{short} in detail here, for dynamical context we note that the system contains a $1.1~M_\odot$ and $0.9~M_\odot$ binary ($\alpha~Cen~A$ and $B$, respectively: spectral types G2V, KV1 \citep{morel_2000}) 
which has a highly eccentric orbit with a semi-major axis of $23.3~$au and an orbital period of $80~$yr \citep{cuello_2024}. The third star in the system, and the closest stellar neighbour to the Sun, is Proxima Centauri, a red dwarf orbiting the pair around $8200~$au ($0.12~M_\odot$; M5.5V)
\citep{kervella_proxima_2017, ribas_2017}). 
In 2016, \cite{anglada-escude_2016} reported the discovery of an Earth-sized planet in the habitable zone around Proxima. There have been other reports of planets in the Proxima system, including a possible sub-Earth-sized inner planet and a super-Earth or mini-Neptune \citep{damasso_low-mass_2020, faria_Proxima_subearth_2022, gratton_searching_2020}. The binary system also has reported planetary candidates, but none confirmed \citep{rajpaul_ghost_2015}, although a planetary system is still believed possible \citep{wagner_imaging_2021}.

The triple star system orbits the Milky Way and is currently approaching the Sun, with an expected closest approach in $\sim28,000$ yr. Thus, any material currently leaving that system at low speed would be heading more-or-less towards the Solar System. Broadly speaking, if material is ejected at speeds relative to its source that are much lower than its source system's galactic orbital speed, the material follows a galactic orbit much like that of its parent, but disperses along that path due to the effects of orbital shear \citep{dehnen_tidalribbons_2018, torres_galactic_2019, Zwart2021}. This behaviour is analogous to the formation of cometary meteoroid streams within our Solar System, and which can produce meteor showers at the Earth. 

Next, in Section~\ref{sec:Methods} we outline the methods used for this study; in Section~\ref{sec:Results} we discuss the results; then we turn to Discussion and Conclusions in Sections~\ref{sec:Discussion} and \ref{sec:Conclusion}.

\section{Methods}\label{sec:Methods}
The study performs the numerical integration of particles released from \alphacen{full} over the last 100~Myr to examine the fastest dynamical pathways from that system to our Solar System. The simulations are performed under the conditions described in the following sections.

\subsection{Galactic Model}\label{sec:Model}

This work adopts a simple Galactic model for the Milky Way which includes only the overall time-independent gravitational field. We neglect some known perturbations that affect small particles in particular, such as ISM drag and magnetic forces. As a result, our model is only applicable to particles above a certain size, which will be addressed in more detail later in this section and in the Discussion (Section~\ref{sec:implications}).

Our simulations are conducted in a galactocentric reference frame, with the Galactic Center (GC) positioned at the origin. A right-handed coordinate system has its x-axis extending through the Sun's projected position on the Galactic midplane towards the GC. The z-axis points towards the North Galactic Pole (NGP). The y-axis is perpendicular to the xz-plane, with its positive direction determined by the right-hand rule.

The Galactic potential model we adopt is of \cite{Miyamoto_1975}, a three-component, time-independent, axisymmetric potential. The potential is smooth and the force of gravity between individual stars is neglected. Individual star potentials can be neglected because the relaxation time, $t_{relax}$ - the time for a body's orbit around the galaxy to be significantly perturbed by interactions with individual stars - is known to be very long in the Milky Way,  $t _{relax} \sim10^{7}~$Gyr \citep{binney_tremaine_2008galacticDyn}. Significantly longer than the age of the Galaxy, indicating that the effect can be neglected over the $\sim100$~Myr time scales considered here.

The Galactic gravitational potential is represented by the sum of the three Galactic components:

\begin{equation}
    \Phi = \Phi_b + \Phi_d + \Phi_h.
\end{equation}

Provided in the galactocentric cartesian coordinate system:

\begin{equation}
    \Phi_{b,h} = -\frac{GM_{b,h}}{\sqrt{x^2+y^2+z^2+b^2_{b,h}}},
\end{equation}

\begin{equation}
    \Phi_{d} = -\frac{GM_{d}}{\sqrt{x^2+y^2+(a_d + \sqrt{z^2+b^2_{d}}})}.
\end{equation}

In these equations, $G$ is the universal gravitational constant, $M_i$ is the total mass of the Galactic component, and $a_i$, $b_i$ are the scale lengths reflecting the geometries of the component. The bulge and halo are portrayed as spherically symmetric, therefore only one scale length is required ($b_{b,h}$). The disk is represented as a flattened spheroid, which requires two scale lengths ($a_d$ and $b_d$). The values we used to initialize our simulation are summarized in Table \ref{Tab:Params}.

\subsubsection{Particle sizes}

\cite{Murray_2004} provide a detailed discussion of the influence of galactic magnetic fields, interstellar medium (ISM) drag, and grain destruction on interstellar particles, and all of these effects become increasingly important as the particle size decreases. The applicability of our model to smaller particles depends on the time and distance traveled in the ISM. We estimate that our model reliably describes rapid delivery (occurring over a fraction of a galactic rotation) for particles 100 microns and larger. A notable size-independent effect not included here is the influence of giant molecular clouds (GMC), which can accelerate the dispersion of galactic meteoroid streams. We will examine these effects post-simulation in the discussion (Section \ref{sec:implications}).

\begin{deluxetable*}{lccccccc}
\tablecaption{The adopted values used to initialize our simulation, including the ICRS coordinates of \alphacen{full} taken from SIMBAD. \label{Tab:Params}}
\tablewidth{0pt}
\tablehead{
\multicolumn{3}{c}{Parameter} & Units & \multicolumn{3}{c}{Value} & References
}
\startdata
\multicolumn{3}{l}{$M_i$, $i=d, b, h$} & $10^{10}M_\odot$ & \multicolumn{3}{c}{7.91, 1.40, 69.80} & [3] \\ 
\multicolumn{3}{l}{$a_i$, $i=d, b, h$} & pc & \multicolumn{3}{c}{3500, 0, 0} & [3] \\
\multicolumn{3}{l}{$b_i$, $i=d, b, h$} & pc & \multicolumn{3}{c}{250, 350, 24000} & [3] \\
\multicolumn{3}{l}{$r_\odot$} & kpc & \multicolumn{3}{c}{8.33$\pm 0.35$} & [4] \\
\multicolumn{3}{l}{$z_\odot$} & pc & \multicolumn{3}{c}{27$\pm 4$} & [2] \\
\multicolumn{3}{l}{$v_\odot$ $(U, V, W)$} & km s$^{-1}$ & \multicolumn{3}{c}{($11.1^{+0.69}_{-0.75}$, $12.24^{+0.47}_{-0.47}$, $7.25^{+0.37}_{-0.36}$)} & [6] \\
\multicolumn{3}{l}{$v_{\odot,circ}$} & km s$^{-1}$ & \multicolumn{3}{c}{218$\pm 6$} & [1] \\
\multicolumn{3}{l}{Galactic Centre Equatorial Coordinates ($\alpha$, $\delta$)} & (hr:min:s, deg/'/") & \multicolumn{3}{c}{(17:45:37.224, -28:56:10.23)} & [5] \\ \hline
\multicolumn{8}{c}{\alphacen{full}} \\ \hline
Star System & $\alpha$ & $\mu_{\alpha}$ & $\delta$ & $\mu_{\delta}$ & Parallax & $v_r$  & References \\ 
 & (deg) & (mas yr$^{-1}$) & (deg) & (mas yr$^{-1}$) & (mas) & (km s$^{-1}$) & \\ 
\hline 
* alf Cen  & 219.873833 & -3608 & -60.83222194 & 686 & 742 & -22.3 & [7] \\
\enddata
\tablereferences{ [1] \cite{Bovy_2015}; [2] \cite{chen2001}; [3] \cite{Dauphole_1995}; [4] \cite{Gillessen_2009}; [5] \cite{Reid_2004}; [6] \cite{schonrich_2010}; [7] \cite{SIMBAD}}
\end{deluxetable*}

\subsection{Simulation}\label{sec:Sim}

Our simulations are based on the Runge-Kutta-Fehlberg method \citep{fehlberg1974}, a fourth-order integrator with variable time step, with an adopted error tolerance of $10^{-6}$. 
The barycenter of \alphacen{short} is initialized into the simulation, as well as a particle representing our Sun (Table~\ref{Tab:Params}). 

We first simulate \alphacen{short} and the Sun backward from $t=0$ (the current epoch) a total of $100~$Myr, slightly less than half of their galactic periods.
From their positions and velocities at $t\approx-100~$Myr, we allow the systems to progress forward $110~$Myr, ending $t\approx10~$Myr in the future, with a time step of $\sim5000~$yr. 
As we propagate the systems forward again, their positions return to within the tolerance threshold at $t=0$, the largest deviation observed being $7\times10^{-5}~$au. As simulation time progresses, an additional 10,000 particles representing ejected material from \alphacen{short} are added every $1~$Myr. These particles play the role of both macroscopic telescopically observable km-class bodies such as asteroids and comets, as well as smaller (sub-mm to m) sized particles that could be detected as meteors in the Earth's atmosphere.

\subsubsection{Ejection model}
Here we will assume that the particles ejected from \alphacen{short} leave the system with some excess velocity, but we will not model their ejection in detail. Instead, we adopt a velocity distribution representative of the ejection process plausibly at work in \alphacen{short} at the present time. Particles could be released by the gravitationally-driven ejection of residual planetesimals by the multiple stars in the system \citep{Bailer-Jones_2018, Cuk_2018, smullen2016_binaries, jackson2018}, or by known or unknown planets \citep{Bailer-Jones_2018, brasser2013, charnoz2003, correa-otto_stability_2019, duncan1987, fernandez1984, Zwart2018}. 

\begin{figure}
   \centering
   \includegraphics[width=\textwidth]{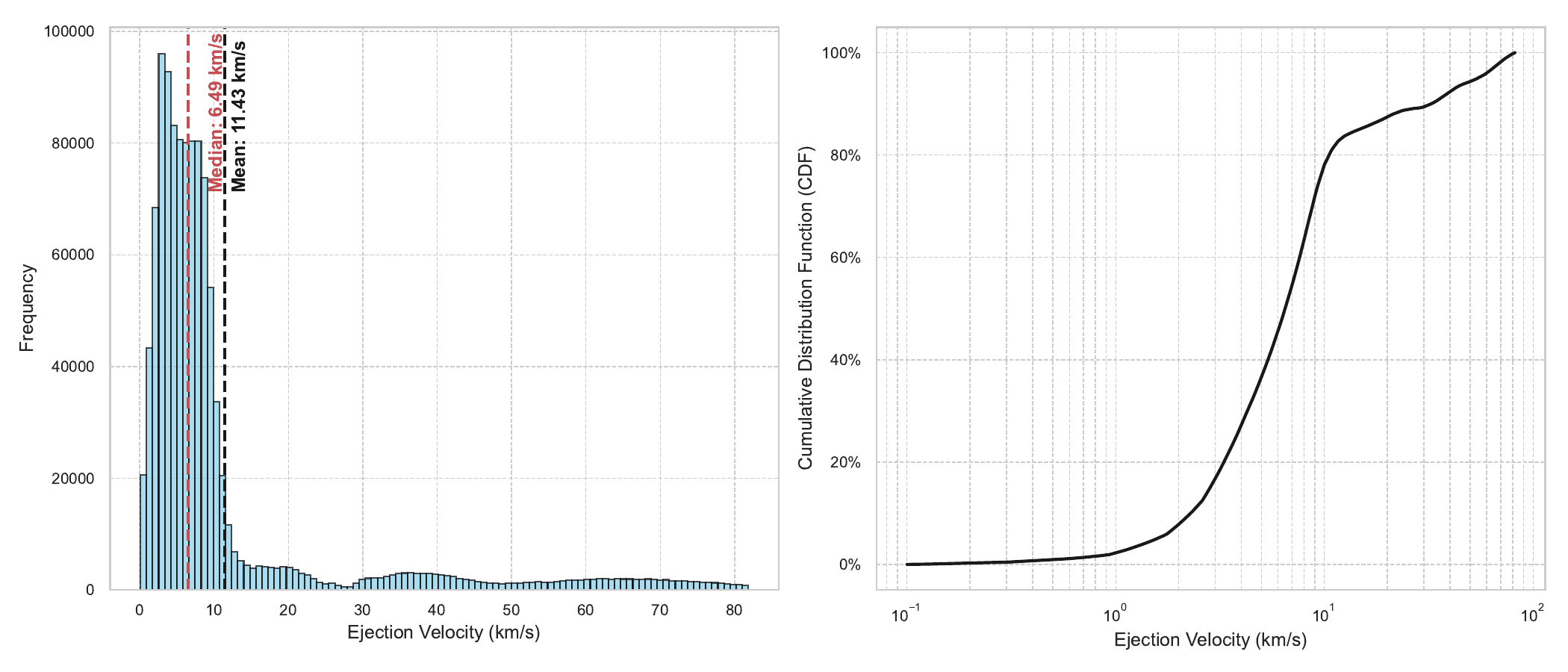}
   \caption{The ejection velocity distribution from our simulations, adapted from Figure 6 of \cite{Bailer-Jones_2018} (their red-dashed line). The right panel is our recreation of the \cite{Bailer-Jones_2018} figure, showing the cumulative distribution of ejection velocities caused from a binary star system with masses $1~M_\odot$ and $0.1~M_\odot$ on a circular orbit of with $10~$au separation.}
   \label{fig:EjectVels}
\end{figure}

\cite{Bailer-Jones_2018} produced an ejection speed distribution for the scattering of planetesimals by a giant planet orbiting a single star, and by a binary star with masses $1~M_\odot$ and $0.1~M_\odot$. Though these parameters do not exactly match \alphacen{short}'s, these models provide a reasonable first approximation to the speed distribution that we might expect for planetary and stellar ejections. We adopt the binary star speed distribution (the red dashed line in Fig. 6 of \cite{Bailer-Jones_2018}) as being most applicable here, and it is shown in Figure \ref{fig:EjectVels}. Note that it includes a low-velocity tail that covers the range of ejection speeds expected for planetary ejections as well,
and indeed we will see in Section \ref{sec:Results} that low-speed ejections are favored in terms of reaching our Solar System.

This ejection speed is taken as the asymptotic speed with which the particles leave \alphacen{short}, $v_\infty$, and is multiplied by a directional vector (chosen randomly from the surface of a unit sphere) and added to the velocity of the star to determine the galactocentric velocity of the ejecta. The new ejected particle is placed within the simulation with the same position as the origin system and with this new velocity vector.

Between each time step of integration, the minimum distance between any particle and the Sun is interpolated by assuming a linear trajectory to avoid missing an encounter by stepping over it. If any particle passes within a specified distance of the Sun (a heliocentric bubble of radius $100,000~$au, chosen more or less arbitrarily but representative of the extent of the outer OC), the object is flagged as a close approach (CA), and all details are noted.

\section{Results}\label{sec:Results}

\begin{figure}
    \begin{interactive}{animation}{AlphaCen_2D.mp4}
    \centering
    \includegraphics[width=0.85\textwidth]{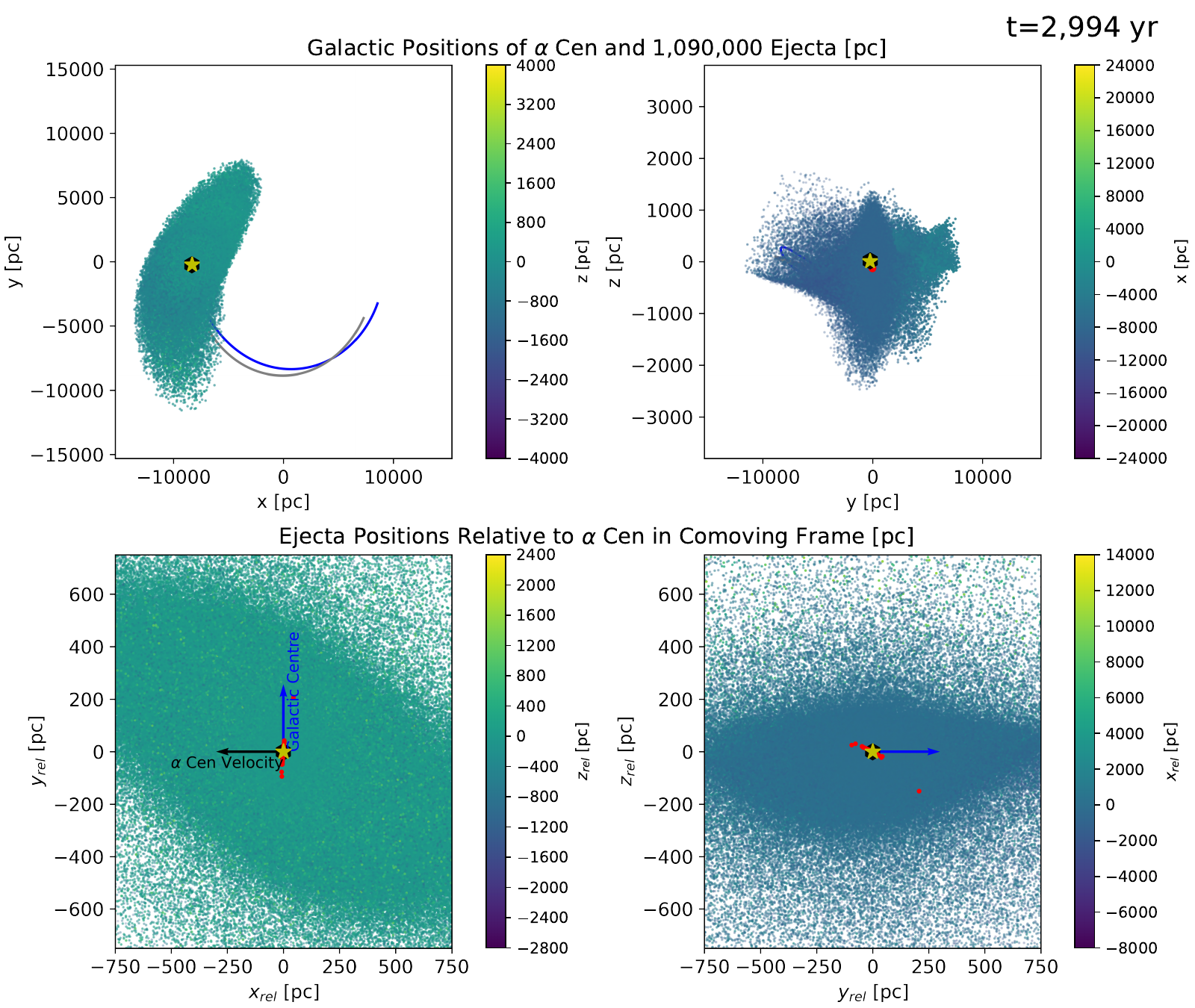}
    \end{interactive}
    \caption{\alphacen{short}'s orbit about the Galactic Centre viewed on the xy and yz planes (top row), as well as the orbits of the ejecta from \alphacen{short} viewed in a comoving frame (bottom row). 
    Our Sun (\textit{Sol}) is marked by a black hexagon and its orbital path indicated by a grey solid line (top row only). \alphacen{short}'s location and path are shown by a yellow star and blue solid line (top row only). In the bottom row,
    the comoving frame follows \alphacen{short} around its orbit while maintaining its orientation with the y-axis pointing towards the Galactic Centre (blue arrow) and \alphacen{short}'s velocity pointing in the -x direction (black arrow). This still frame is taken at $t\approx3,000~$yr (that is, +3,000 years from the current epoch) after $\sim100$ Myr of integration. The colours of the ejecta represent the 3rd dimension of position, except that any particle that will at any point come within $100,000~$au of \textit{Sol} are plotted in red. The full animation is available in the HTML version of this publication which shows the time evolution from $t\approx-100~$Myr to $t\approx10~$Myr. The duration of the animation is 11 s. \url{https://youtu.be/YABoYgNKr-I}}
    \label{fig:ACen_4plot}
\end{figure}

A total of $1.09\times10^6$ particles were ejected from the \alphacen{short} system throughout the simulation. As described in Section \ref{sec:Sim}, the particles were ejected in random directions and their ejection speeds followed the distribution of \cite{Bailer-Jones_2018}.
The simulation began at $t\approx-100~$Myr and ended at $t\approx10~$Myr (Figure \ref{fig:ACen_4plot} and the corresponding animation). 

\begin{figure}
    \begin{interactive}{animation}{AlphaCen_SunCent.mp4}
    \centering
    \includegraphics[width=0.85\textwidth]{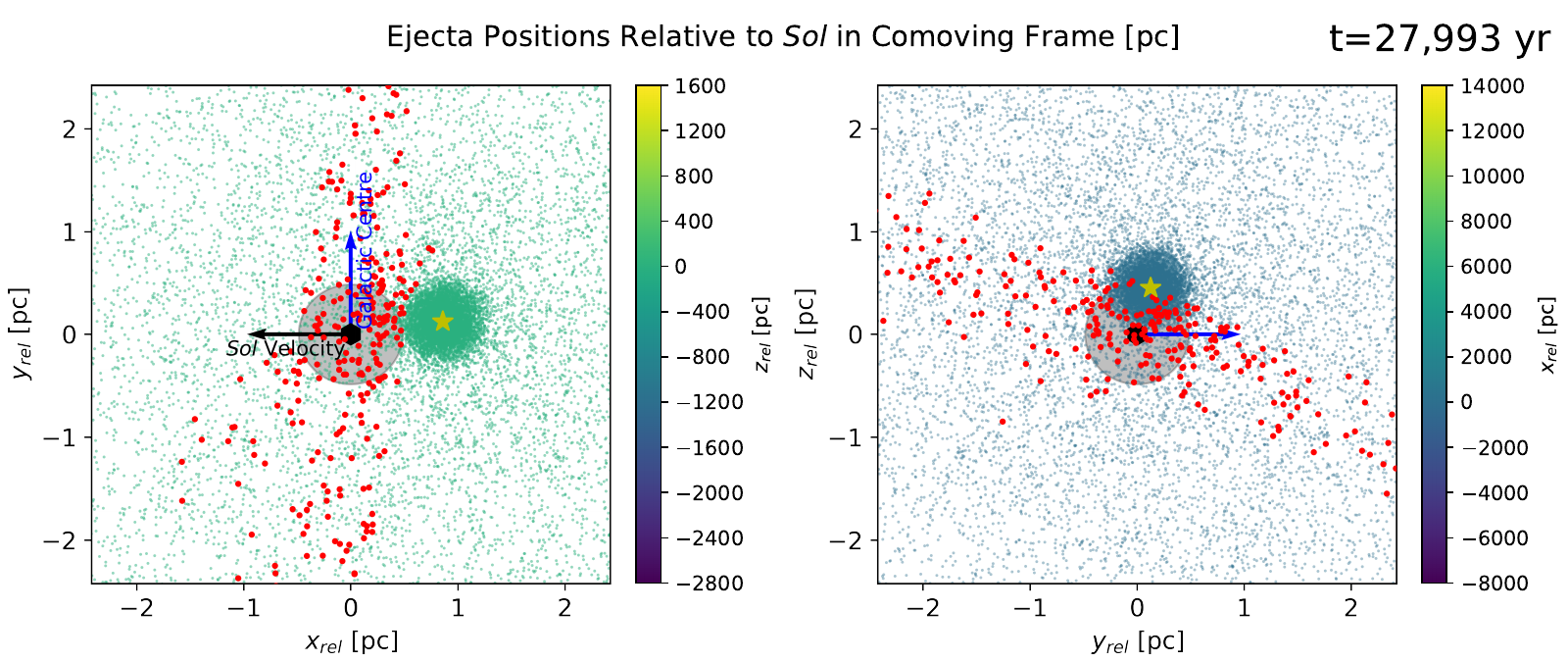}
    \end{interactive}
    \caption{The paths of the ejecta from \alphacen{short} viewed in a comoving frame with \textit{Sol}. The comoving frame follows \textit{Sol} around its orbit maintaining the orientation with the y-axis pointing towards the Galactic Centre (blue arrow) and \textit{Sol}'s velocity pointing in the -x direction (black arrow). Our Sun (\textit{Sol}) is indicated by a black hexagon and \alphacen{short} by a yellow star. This still frame is taken at closest approach to our Solar System ($t\approx28,000~$yr, years from the current epoch) after $\sim100$ Myr of integration. The colours of the ejecta represent the 3rd dimension of position. The grey circle represents the extent of the Oort cloud ($100,000~$au), any particle that comes within this distance of \textit{Sol} at any point is flagged as a close approach. These particles are plotted in red. The full animation is available in the HTML version of this publication which shows the shower duration from $t\approx-3~$Myr to $t\approx7~$Myr. The duration of the animation is 44 s. \url{https://youtu.be/qxd5kguXRWw}}
    \label{fig:ACen_sunCent}
\end{figure}

\begin{figure}
    \centering
    \includegraphics[width=0.45\textwidth]{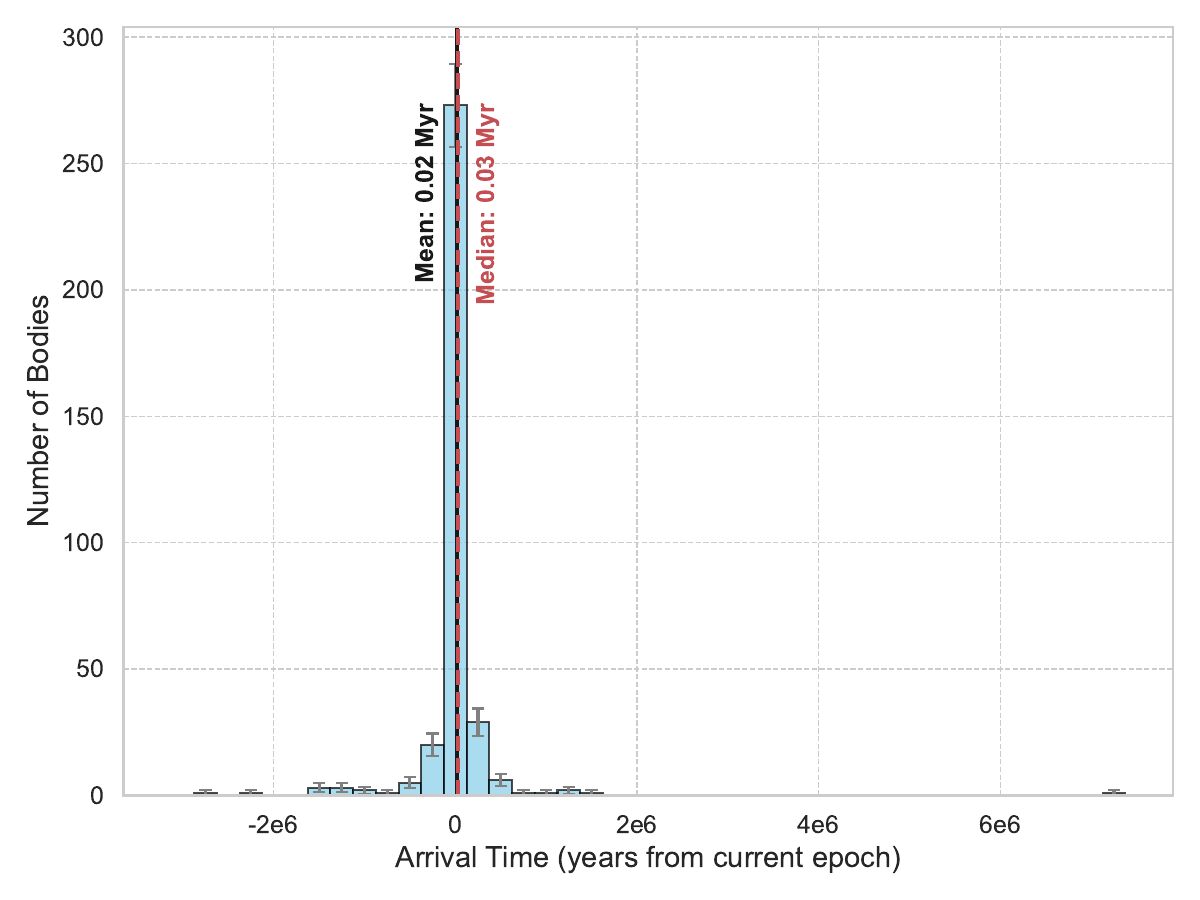}
    \includegraphics[width=0.45\textwidth]{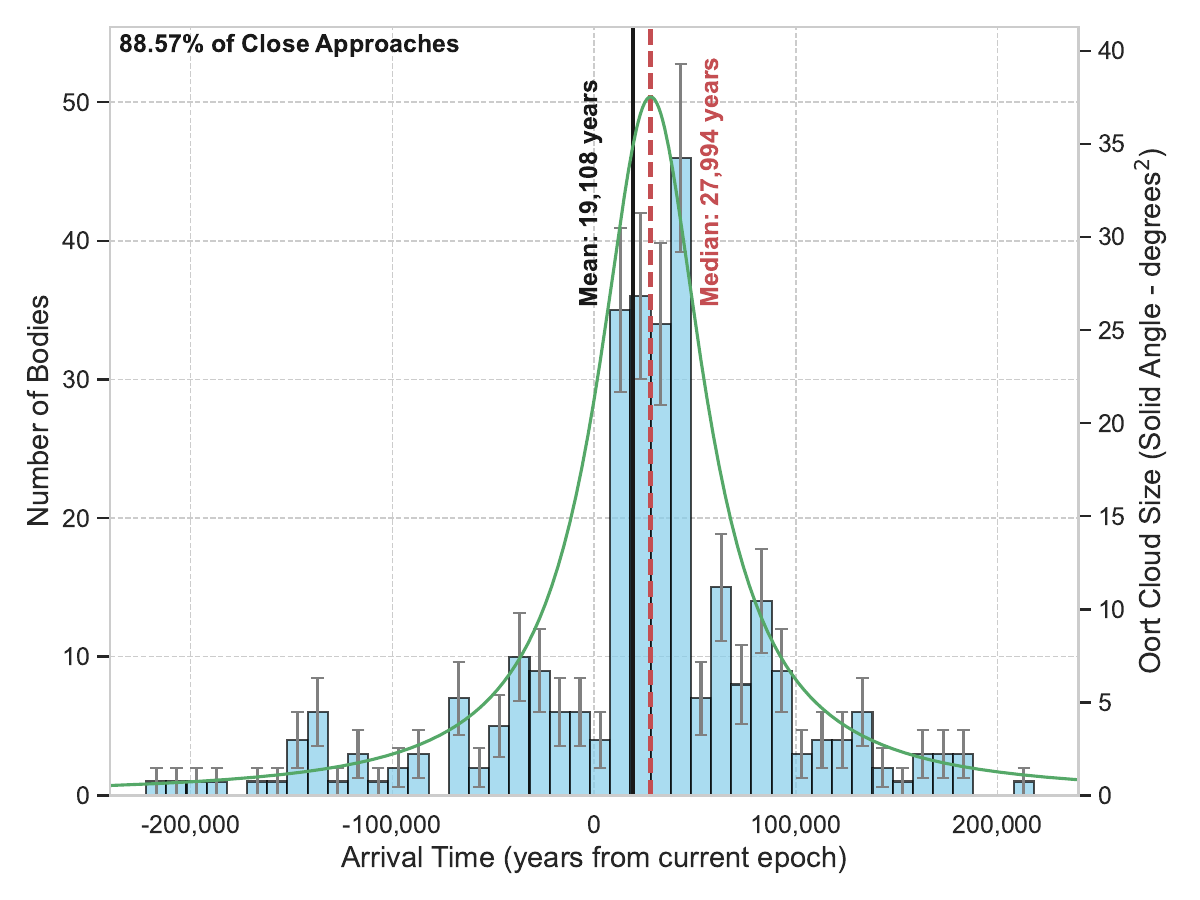}
    \caption{The arrival times at the Solar System of the \alphacen{short} material. The left figure shows all close approaches while the right zooms into the time of peak intensity. The green line shows the effective cross-section of the Solar System (the solid angle subtended by our Oort cloud as seen from \alphacen{short}).
    }
    \label{fig:alphaCen_CA_ArrivalTimes}
\end{figure}

Only a small fraction of the \alphacen{short} ejecta come within the CA distance of the Sun. In total, 350 particles had a CA with the Solar System, $\sim$0.03\% of the total ejecta. The first CA arrives at $t\approx-2.85~$Myr. Material continues to arrive for $\sim10~$Myr (Figure \ref{fig:ACen_sunCent} and the corresponding animation) with the majority of CAs within $\pm200,000~$ yr of the current epoch (Figure \ref{fig:alphaCen_CA_ArrivalTimes}). This peak is centered around \alphacen{short}'s time of closest approach to the Solar System in $\sim28,000$ yr. In that figure we also include the effective cross-section of the Solar System (here taken to be of $10^5$~au radius) as viewed from \alphacen{short} and expressed as a solid angle. This cone reaches a maximum full-width of $37.5^\circ$ when \alphacen{short} is at its point of closest approach, and we see that the rate of arrivals broadly coincides with and peaks at the same time. However, because the particle travel times are not zero, this is only an approximation of the effective cross-section that they see and which depends on their relative velocity. Nevertheless, it illustrates that we expect material to be transferred more efficiently as our Solar System's apparent cross-section grows as we move towards our point of closest approach with \alphacen{short}.

\begin{figure}
    \centering
    \includegraphics[width=0.45\textwidth]{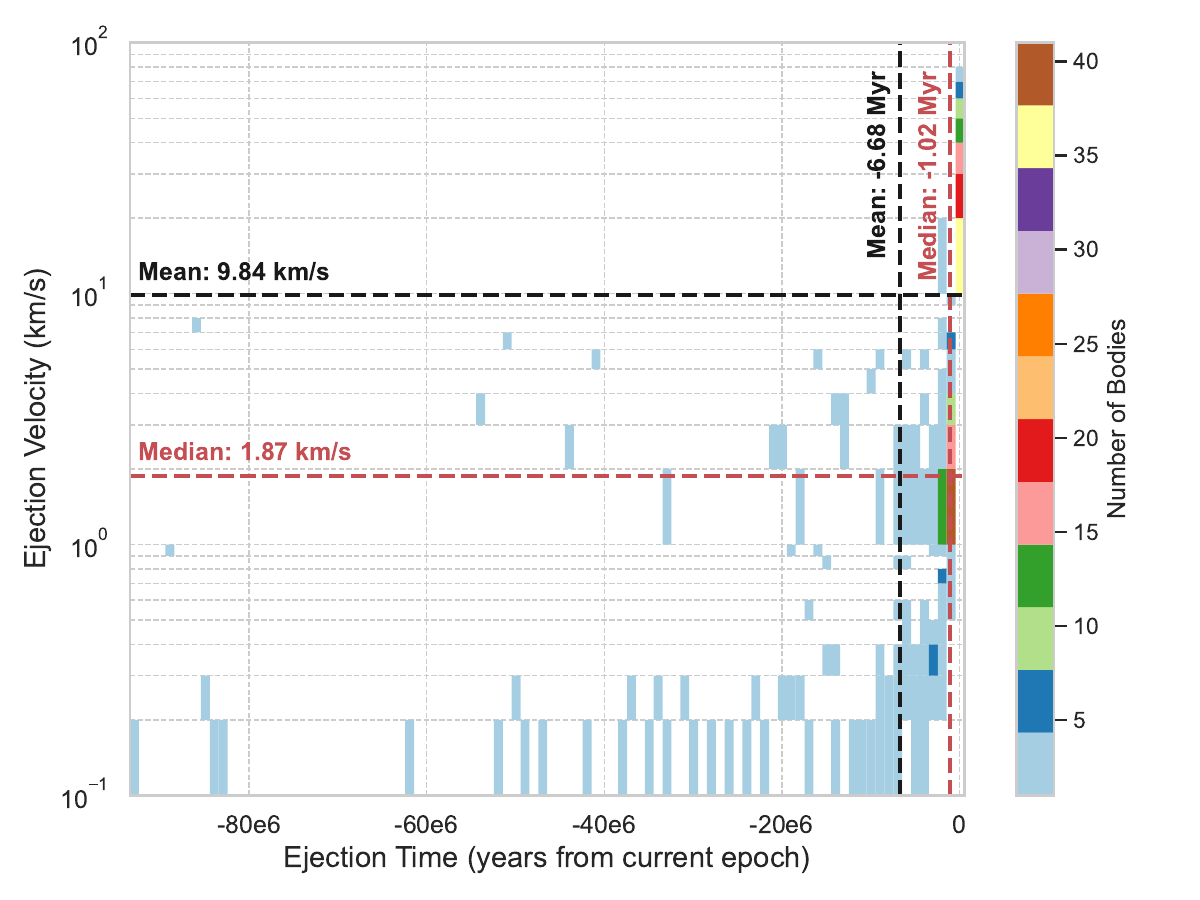}
    \includegraphics[width=0.45\textwidth]{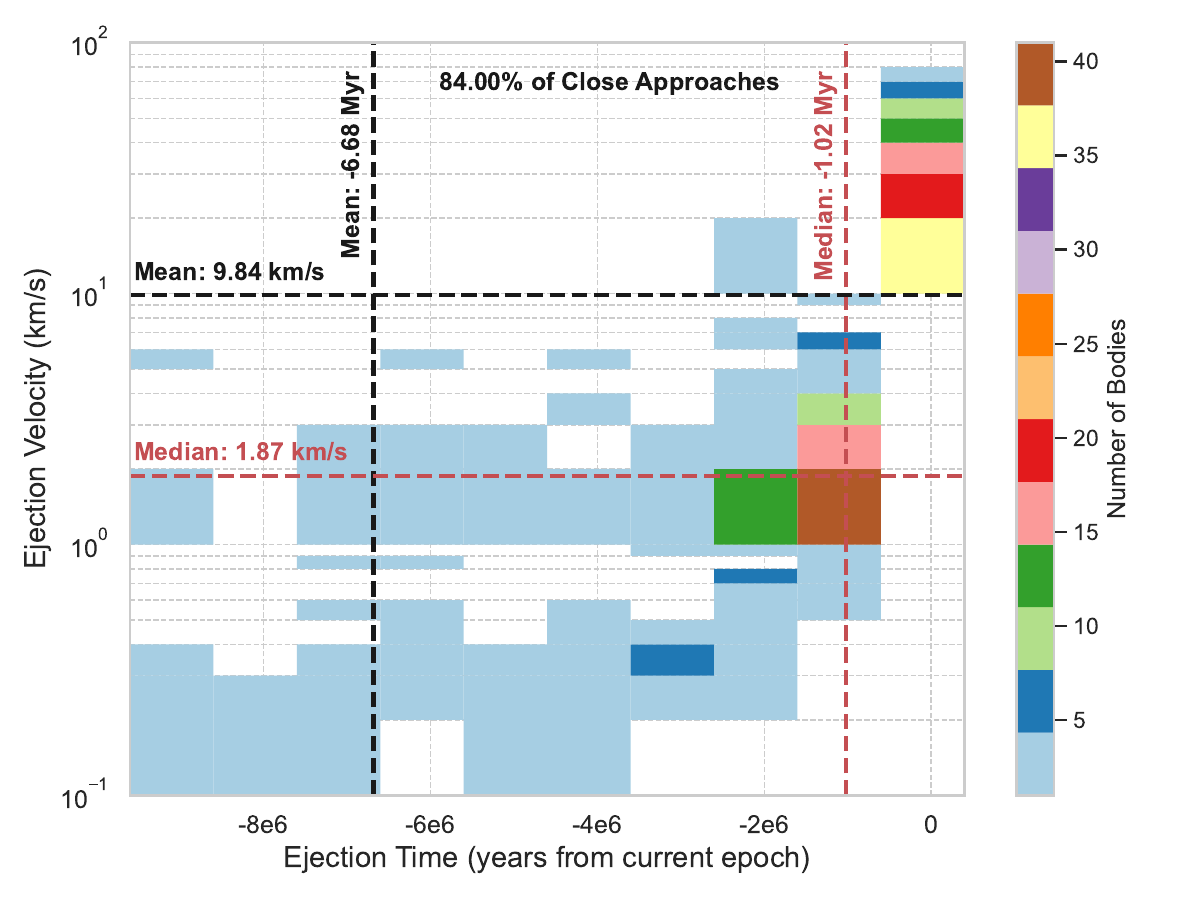}
    \caption{The ejection velocities vs time of ejection of the \alphacen{short} material that enter our Oort cloud. The left figure shows the total distribution while the right zooms into the most common times.}
    \label{fig:alphaCen_CA_EjectionTimeVel}
\end{figure}

The ejection speed distribution from a giant planet peaks between 1 and 2 km s$^{-1}$ \citep{Bailer-Jones_2018}; while in our simulations we find 52\% of particles that arrive from \alphacen{short} have ejection speeds less than 2 km s$^{-1}$ (with a maximum around $77~{\rm km~s}^{-1}$, Figure \ref{fig:alphaCen_CA_EjectionTimeVel}). Thus, the speeds that most effectively move material from \alphacen{short} to us closely match those expected for ejection by a massive planet, such as might occur due to the planets around Proxima Centauri, though no planets more massive than a mini-Neptune have yet been confirmed in the \alphacen{short} system.


\begin{figure}
    \centering
    \includegraphics[width=0.45\textwidth]{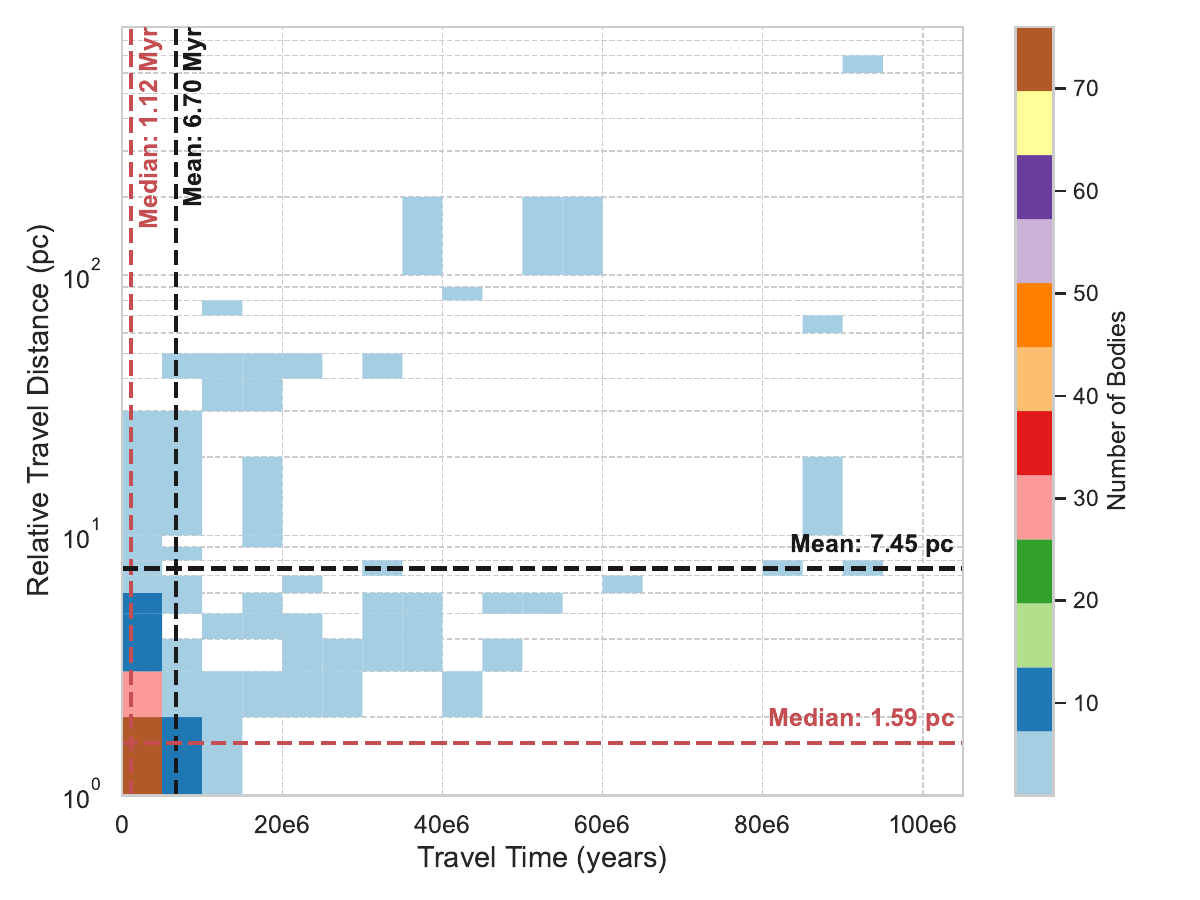}
    \includegraphics[width=0.45\textwidth]{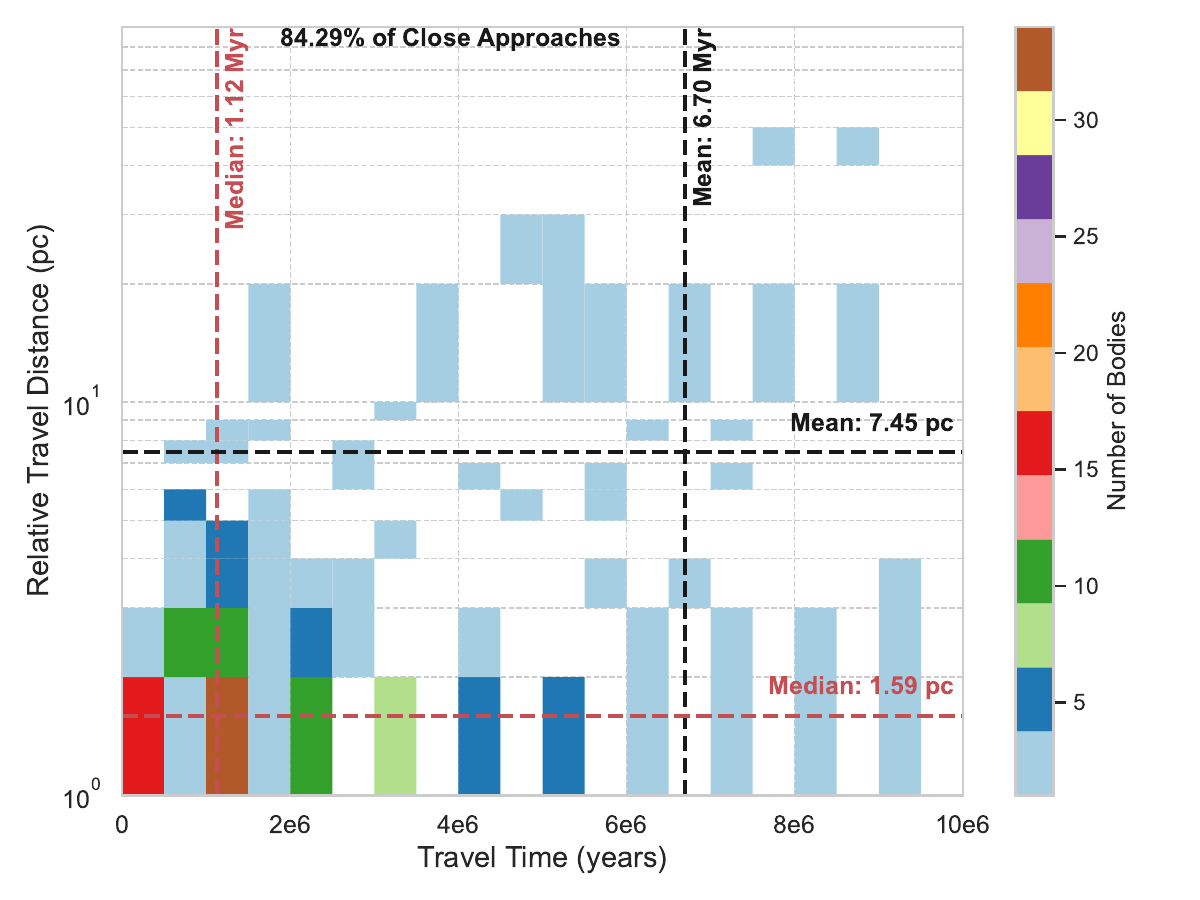}
    \caption{The relative travel distance (the distance a particle travels with respect to \alphacen{short}) vs travel time of the \alphacen{short} material that enter our Oort cloud. The left figure shows the total distribution while the right zooms into the most common times.}
    \label{fig:alphaCen_CA_TravelTimeDist}
\end{figure}

The times of ejection ($t_{eject}$) that resulted in a CA ranged from $t_{eject}\approx-93~$Myr to $t_{eject}\approx-12,000~$yr. 
The majority of the CAs had ejection times closer to the current epoch ($\approx53\%$ with $t_{eject}>-1$ Myr; $\approx84\%$ with $t_{eject}>-10$ Myr), meaning that fresh ejecta was more likely to encounter the Solar System (Figure \ref{fig:alphaCen_CA_EjectionTimeVel}). This is as expected, as the \alphacen{short} system is moving towards the Sun and the solid angle the Solar System subtends is largest when their mutual distance is smallest. The majority ($\approx84\%$) of the CAs traveled for $<10~$Myr, a fraction of a galactic orbit (Figure \ref{fig:alphaCen_CA_TravelTimeDist}). These particles are therefore exposed to the ISM for a relatively short time, lessening any effects from the galactic magnetic field, ISM drag, grain destruction, and perturbations from GMC encounters (discussed further in Section \ref{sec:implications}).

\begin{figure}
    \centering
    \includegraphics[width=0.45\textwidth]{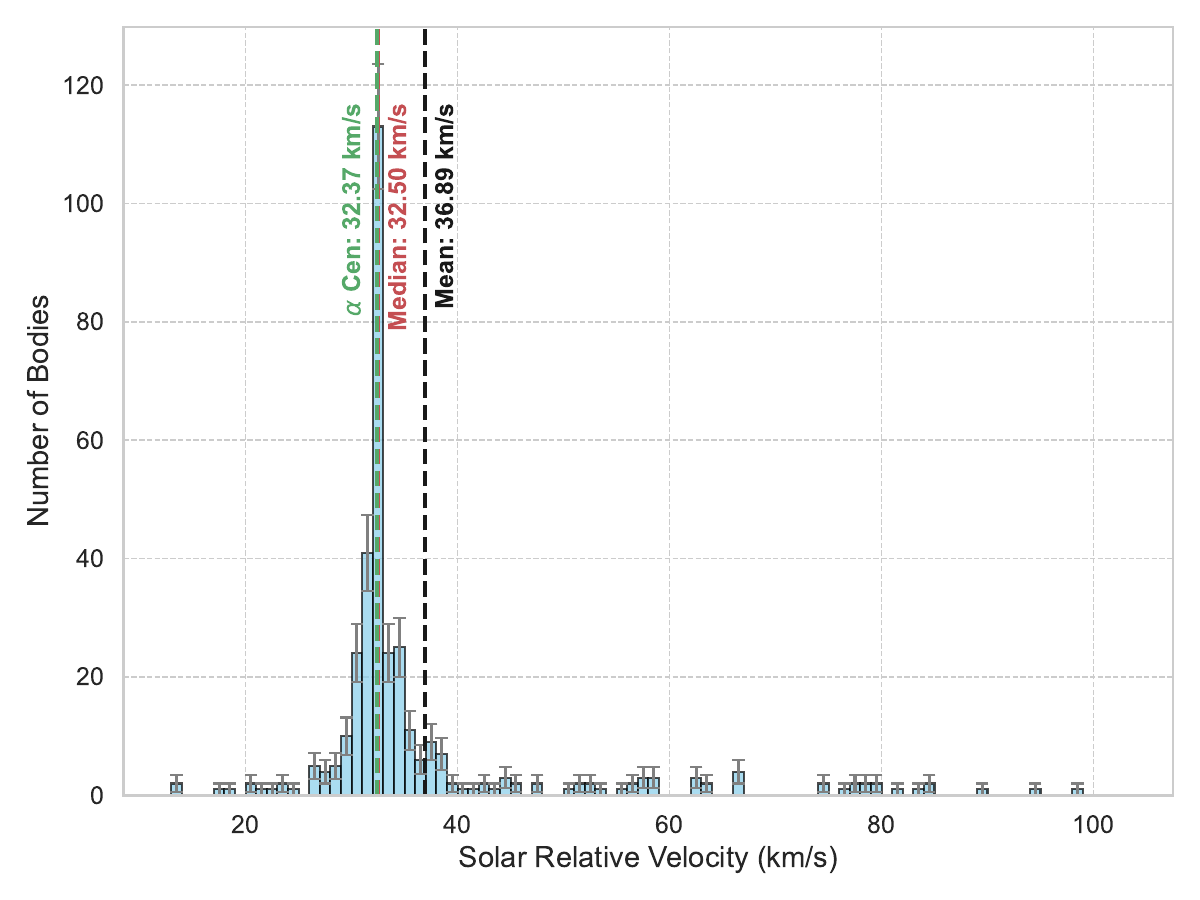}
    \includegraphics[width=0.45\textwidth]{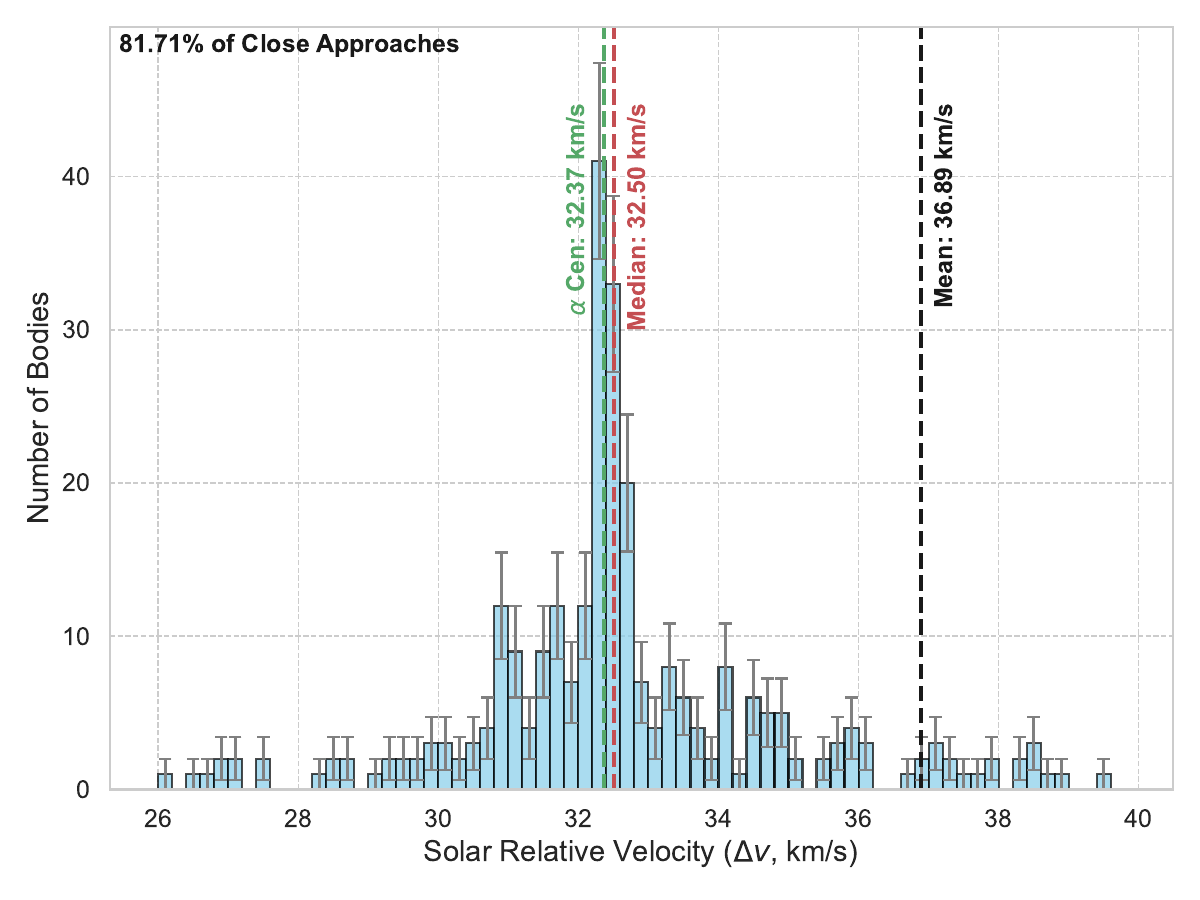}
    \caption{The apparent velocities of the \alphacen{short} material that cross our Oort cloud. The left figure shows the total distribution while the right zooms into the most common speeds.}
    \label{fig:alphaCen_CA_SolarRelVel}
\end{figure}

When the CAs intercept the Solar System, their median apparent velocity relative to the Sun ($\Delta v$) observed at their closest approach is $32.50~{\rm km~s}^{-1}$, ranging from $13.75~{\rm km~s}^{-1}$ to $103.17~{\rm km~s}^{-1}$ (Figure \ref{fig:alphaCen_CA_SolarRelVel}). Due to the high fraction of CAs that resulted from low ejection speeds, we see the solar relative velocities center around the current apparent velocity of \alphacen{short} ($\Delta v=32.37~{\rm km~s}^{-1}$) as expected. 

As particles fall into the Solar System, their heliocentric speed $v_{hel}$ increases beyond the relative speed $\Delta v$ that they have at large distances. If they are observed on Earth as meteors, their observed heliocentric speed will be given by

\begin{equation}
 v_{hel} = \sqrt{\Delta v^2 + \frac{2GM_{\odot}}{r_{\oplus}}}  
\end{equation}
where $\Delta v$ is its velocity relative to the Sun when distant from our system and $r_{\oplus}$ is the Earth's distance from the Sun.
Using our minimum, median, and maximum values found, we can determine an expected range of heliocentric speeds in the inner Solar System, at $1~$au from Sun, to be $44\lesssim v_{hel} \lesssim 111~{\rm km~s}^{-1}$ with a median around $53~{\rm km~s}^{-1}$.

\subsection{Meteor Radiants}\label{sec:Radiants}

\begin{figure}
    \centering
    \includegraphics[width=0.45\textwidth]{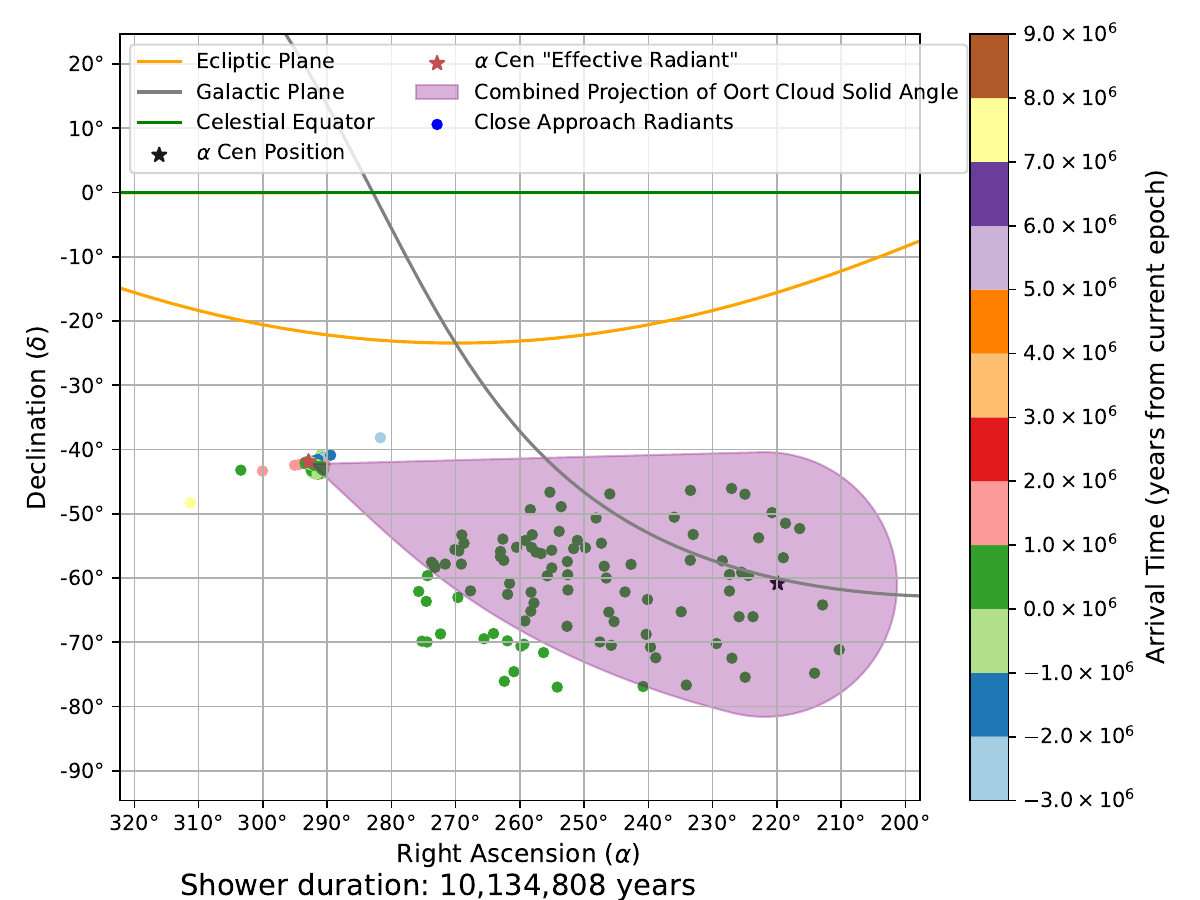}
    \includegraphics[width=0.45\textwidth]{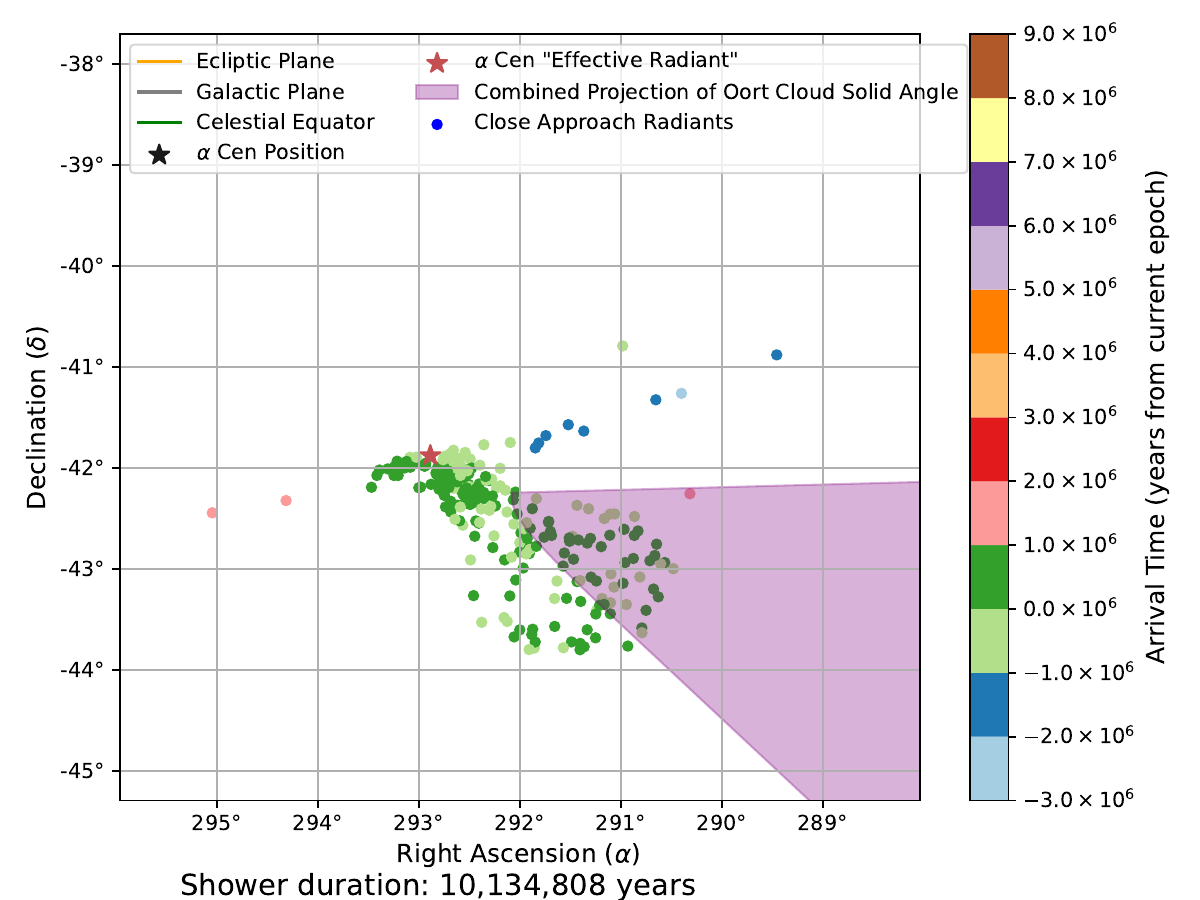}
    \caption{The heliocentric equatorial radiant for the 350 close approaches at the time of their closest Solar approach (``Arrival Time"), with the current heliocentric equatorial coordinates of \alphacen{short} plotted as a black star and the ``effective radiant" corresponding to \alphacen{short}'s apparent velocity is plotted as a red star. The purple shaded region is the combined projection of the the effective cross-section of the Solar System (solid angle size as seen from \alphacen{short}), from 
    the start of the simulation up to the current time ($t\approx0~$yr).
    The left figure views the entire CA population while the right figure zooms into the region surrounding \alphacen{short}'s effective radiant.
    }
    \label{fig:alphaCen_CA_radiantZoom}
\end{figure}

\begin{figure}
    \begin{interactive}{animation}{PeakAnimation_ICRS.mp4}
    \centering
    \includegraphics[width=0.85\textwidth]{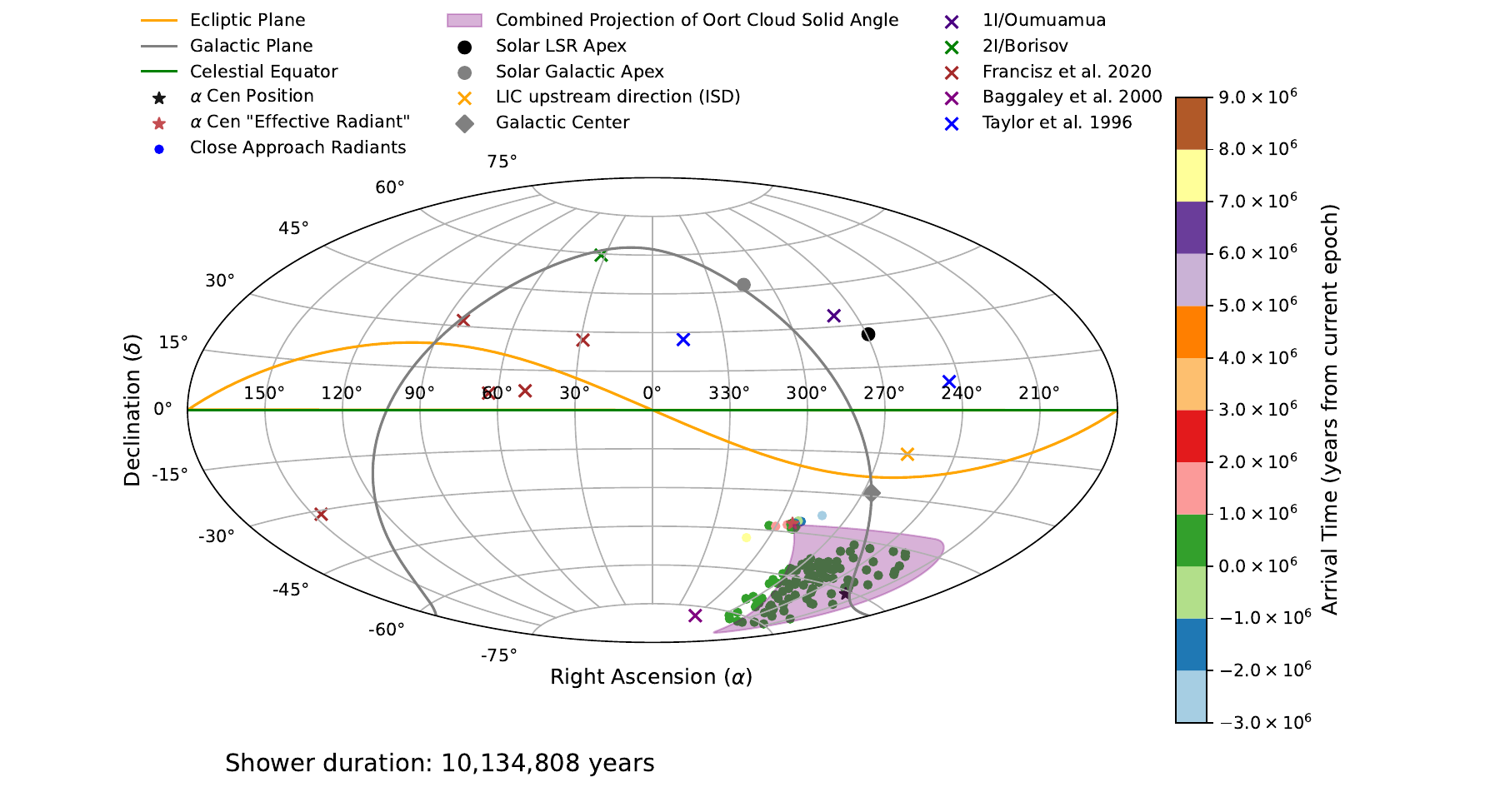}
    \end{interactive}
    \caption{
    An all-sky version of Figure \ref{fig:alphaCen_CA_radiantZoom} (see that Figure for additional information), with the addition of the two confirmed km-scale interstellar objects (ISOs), 5 unconfirmed interstellar meteors, 3 suggested high interstellar meteor flux regions, and the direction of the local interstellar cloud (LIC) upstream direction of interstellar dust (ISD) for comparison (all summarized in Table \ref{tab:ISO}). The Solar apex with respect to the Local Standard of Rest (LSR) and within the Milky Way are also included, along with the direction towards the Galactic Center.
    In the full HTML version of this publication an animation is available showing the radiants of the CAs as they appear during peak intensity, and the progression of \alphacen{short}'s position and ``effective radiant." The duration of the animation is 41 s. \url{https://youtu.be/bu6849CAHkg}}
    \label{fig:alphaCen_CA_radiantAnimation}
\end{figure}

The heliocentric equatorial radiant of CAs is plotted in Figure \ref{fig:alphaCen_CA_radiantZoom}, and also plotted and animated for the  resulting ``interstellar meteor shower" in Figure \ref{fig:alphaCen_CA_radiantAnimation}. The radiant would be the on-sky location from which the interstellar particle would appear to originate if it appeared as a meteor in Earth's sky. The figure converts the solar relative velocity ($\Delta v$) to a heliocentric equatorial right ascension (RA or $\alpha$) and declination (Dec or $\delta$) for easy comparison with Solar System meteor shower radiants. Two radiant clusters appeared in our simulations, one centered on an average position of ($\alpha$, $\delta$)=($292 \pm 1^\circ$, $-43 \pm 2^\circ$) and the other at ($\alpha$, $\delta$)=($249 \pm 17^\circ$, $-60 \pm 8^\circ$). 

The first cluster is located at the on-sky position of what we term the ``effective radiant", the projection of the \alphacen{short}'s velocity vector relative to the Sun (reversed) onto the celestial sphere. Currently \alphacen{short}'s motion relative to the Sun is one primarily of approach along the radial direction. Therefore particles originating from it and that arrive here must be traveling along that nearly same line, and thus the effective radiant coincides largely with \alphacen{short}'s position on the sky. One would expect the radiant of the stream to be in the approximate part of the sky as the origin star as it is approaching the Solar System. In the animation linked to Figure \ref{fig:alphaCen_CA_radiantAnimation}, we do indeed see the majority of the CAs coming from the same direction as both the on-sky position of \alphacen{short} and its effective radiant, at least at times before the current epoch. 

As \alphacen{short} passes its closest point to us, its motion changes from being mostly radial to mostly tangential. Though particles continue to arrive in the Solar System from the direction of the effective radiant, \alphacen{short} itself moves away from this point on the sky. The apparent cross-section of our Solar System is largest at closest approach, allowing material ejected with a range of speeds within a much wider cone to reach us, creating the broad second radiant cluster.

After \alphacen{short} passes its closest approach, it begins to recede from us at $32~{\rm km~s}^{-1}$; however, low-speed material ejected in the past would still arrive from the same effective radiant. But at this point it is difficult for material newly ejected from that system to reach us, and the rate of arrival of material from \alphacen{short} drops sharply.

The effective cross-section of the Solar System shows a smooth progression up to the peak intensity of the \alphacen{short} shower (Figure \ref{fig:alphaCen_CA_ArrivalTimes}). In our simulation, we instead see a more sudden increase, which result from our discrete ejection modeling. As we eject material every $1~$Myr, we see a sudden burst of CAs around the time of \alphacen{short}'s closest approach, and a sudden shift in the radiant, rather than a gradual change. To examine this in more details,  we ran an auxiliary simulation just during closest approach where the time between ejections was shortened. In this case, the CA radiants are smoothly dispersed along the path of the projected cross-section of the Solar System (purple shaded region in Figures \ref{fig:alphaCen_CA_radiantZoom} \& \ref{fig:alphaCen_CA_radiantAnimation}) from a narrow cluster surrounding \alphacen{short}'s effective radiant to the large region near \alphacen{short}'s current on-sky position.

\subsection{Are There \texorpdfstring{$\alpha~Cen~$}~Particles among proposed interstellars?}

A selection of previously discovered/proposed interstellars are plotted along side the observed radiants of our \alphacen{short} shower in Figure \ref{fig:alphaCen_CA_radiantAnimation}. Included are: two confirmed macroscopic interstellar objects (1I/'Oumuamua and 2I/Borisov), traced back to their asymptotic arrival directions \citep{Hallat_2020}; five unconfirmed interstellar meteoroid candidates observed by the Canadian Meteor Orbit Radar (CMOR) \cite{Froncisz_2020}); three reported interstellar meteor influx directions detected by the Advanced Meteor Orbit Radar (AMOR) \citep{baggaley2000, taylor_discovery_1996}; along with confirmed interstellar dust (ISD) influx directions into the Solar System as reported by spacecraft Cassini, Ulysses, and Galileo \citep{sterken_flow_2012}. Other directions are also included for comparison, including the Solar apex with respect to the Local Standard of Rest (LSR) \citep{jaschek_solar_1992}; the Solar apex with respect to the Galactic Center (Solar Galactic Apex) \citep{Reid_2004}.

The list of values plotted in Figure~\ref{fig:alphaCen_CA_radiantAnimation} is provided in Table \ref{tab:ISO}. None of these appears to be correlated with the directions expected for \alphacen{short}. Further analysis of meteor databases to search for \alphacen{short}-related events is encouraged.

\begin{deluxetable*}{lcccc}
\tablecaption{Reported interstellars, including LIC upstream direction and significant directional values in ICRS coordinates.\label{tab:ISO}}
\tablewidth{0pt}
\tablehead{
Designation & Right Ascension ($\alpha$) & Declination ($\delta$) & Category & References \\
         & (deg) & (deg) &  &  
}
\startdata
1I/'Oumuamua & 279.6 & 33.9 & ISO & [3] \\
        2I/Borisov   & 32.6 & 59.5 & ISO & [3] \\
         & 29.1  & 26.8 & meteor & [2] \\
         & 49.6  & 7.2 & meteor & [2] \\
         & 63.7 & 6.2 & meteor & [2] \\
         & 82.6 & 32.0 & meteor & [2] \\
         & 146.6 & -31.1 & meteor & [2] \\
         & 299.0 & -78.4  & meteor flux & [1]  \\
         & 244.0 & 9.2 & meteor flux & [6] \\
         & 347.1 & 27.3 & meteor flux & [6] \\
         Local interstellar cloud (LIC) upstream direction& 259 & 8 & ISD & [5] \\
         \hline \\
         Solar apex relative to Local Standard of Rest (LSR) & 258.7 & -15.0 &  & [4] \\
         Solar Galactic Apex & 313.3 & 47.5 &  & This work \\
        Galactic Centre (GC) -  ($\lambda$,$b$)=($0^{\circ}$,$0^{\circ}$) & 266.4 & -28.9 & & [7] \\
\enddata
\tablereferences{[1] \cite{baggaley2000},
[2] \cite{Froncisz_2020}, 
    [3] \cite{Hallat_2020},
    [4] \cite{jaschek_solar_1992},
    [5] \cite{sterken_flow_2012},
    [6] \cite{taylor_discovery_1996},
    [7] \cite{Reid_2004}
    }
\end{deluxetable*}

\section{Discussion}\label{sec:Discussion}

Earlier sections have shown that there are plausible dynamical pathways from \alphacen{short} to our Solar System. We now turn to a discussion of the particle sizes that can survive the passage from \alphacen{short} to our Solar System, and the expected flux of such particles.

\subsection{Grain sizes \& Implications}\label{sec:implications}

Small particles traveling through the interstellar medium (ISM)
are subject to a number of effects not modeled here. Thus, though in principle particles may travel between \alphacen{short} and our Solar System, whether or not particles --in particular mm-sized and smaller particles that might be observed as meteors in Earth's atmosphere-- can survive the journey depends on various factors, such as travel time and speed relative to the ISM. 
Following \cite{Murray_2004}, we can compute the minimum particle size that can survive the journey from \alphacen{short} to our Solar System. In particular, they examine three effects 1) whether the effect of magnetic fields on small charged grains will deflect them significantly, 2) whether drag against the ISM will halt the grains and 3) whether the grain will be destroyed through sputtering by high-speed gas atoms or by grain-grain collisions.

We extracted the relevant parameters for each of the 350 CAs from our simulation and computed the minimum size needed for a grain traveling along that trajectory to survive all three effects (\cite{Murray_2004}'s equations 44, 45, and 47 relating to drag force, gyroradius, and grain destruction, respectively).
We find that minimum particle sizes ($a$, radius) of 
$1.05\leq a\leq73.22~{\rm \mu m}$, with a median of $3.30~{\rm \mu m}$, can survive their journey (using values: $\rho=3.5~{\rm g~cm}^{-3}$; $n_H=1~{\rm cm}^{-3}$; $U=1~$V, $B=5~{\rm \mu G}$; and $T_{gas}=10^6~$K). To clarify, our median particle size corresponds to a simulated particle that traveled 1.5 pc in \alphacen{short}'s comoving frame at an average velocity of 16 km s$^{-1}$ with respect to the circular velocity of the Sun. At this size and speed, the particle can travel 125 pc in the ISM before grain destruction becomes relevant, 4200 pc for ISM drag, and only 1.5 pc for magnetic forces, and thus our typical particles are effectively magnetically limited. In fact, all of our particles are limited by magnetic forces.

At these small grain sizes, detectability by meteor radar instruments is limited. The practical lower limit to the size of meteoroids detected by currently operating meteor patrol radars like the Canadian Meteor Orbital Radar (CMOR, described further in the next section) is roughly 100 $\mu$m in diameter. Particles at these sizes travel unimpeded for large distances across the Galaxy. For example, a 100 $\mu$m particle travelling at our typical speed of 17~km~s$^{-1}$ could travel 369 pc before being substantially deflected by magnetic fields. Only 1 of our simulated particles traveled farther (600 pc), while the remaining 349 traveled $<200$ pc (Figure \ref{fig:alphaCen_CA_TravelTimeDist}). 
We therefore adopt a fiducial meteor size of $100~{\rm \mu m}$ (diameter) at which the effects of drag, grain destruction and magnetic deflection do not present a concern in this context.
But in any case, from this analysis we can conclude that our results are applicable from submillimeter particles up to large telescopically observable asteroids or comets.

We note an important size-independent effect that we ignore is the effect of encounters with GMCs, which can create $\sim$10 km s$^{-1}$ kicks in velocity on objects orbiting with the Milky Way's disk, and which occur with a characteristic time of 200 Myr \citep{wielen1977diffusion,mihalas1981galactic}. Our simulations here only extend for 100~Myr and most of the mass transfer observed occurs over much shorter time scales, so the effect of GMCs on our results is likely negligible.

\subsection{Mass/Number Influx Estimates}\label{sec:MassDensity}

We can now begin to draw some conclusions on how many \alphacen{short} particles we may expect to see in the Solar System currently and during the peak intensity of the shower. To do this, we must estimate the ejection rate for the \alphacen{short} system. 

Unfortunately, the rate of ejection of material from \alphacen{short} is poorly constrained. As a first-order approximation, we assume that \alphacen{short} ejects material at a rate similar to that of the Solar System at the current time. At macroscopic sizes, this includes ejection of asteroids recently escaped from the asteroid belt, as well as the ejection of comets arriving from the OC or stripped from it by the galactic tidal field. The best quantified of these processes is the ejection of new OC comets, and we will adopt it here as a proxy for our total ejection rate. Comets newly arriving in the inner Solar System from the OC are on nearly-unbound orbits, and planetary perturbations will scatter half onto more tightly bound orbits and eject the other half \citep{wei79, wietre99}. So, the rate at which OC comets are ejected in our Solar System is roughly half the arrival rate of new OC comets to the inner Solar System. The rates of long-period comet detection have been carefully studied. \cite{Boe2019} report a size-frequency distribution (SFD) of long-period comets, which are mostly new OC comets, based on observations from Pan-STARRS. During the 6.8 year survey in question, 229 comets were observed with a typical nucleus size of 4~km.
Extending this to smaller but still telescopically accessible sizes using their observed size distribution, this corresponds to $1.76 \times 10^3$ new OC objects of $100~$m or greater diameter passing perihelion per year, or an ejection rate of $\sim 0.5 \times1.76\times10^{3} = 9\times10^{2}$ objects $>100$~m per year. Therefore, if \alphacen{short} ejects cometary material at a rate equal to our own Solar System, then  $9 \times 10^8$ comets are ejected per Myr. In our simulations, $10^4$ particles are ejected every Myr, so a single simulated particle represents $f_{sim} = 9\times 10^8/ 10^4 = 9 \times 10^4$ real particles $> 100$~m in diameter. 

In our simulations, at the current time, approximately 5 particles are seen to enter the OC per $10,000$ yr (Figure~\ref{fig:alphaCen_CA_ArrivalTimes}). Since each of our simulated particles corresponds to $f_{sim}$ real objects larger than $100$~m, this translates into a flux into the OC of 
$5 f_{sim}/10^4 = 45$ macroscopic particles per year.

Particles from \alphacen{short} arrive with typical speeds of $32~{\rm km~s}^{-1}$ relative to the Sun and so take an average of $\sim20,000$ years to cross the 200,000~au diameter of the OC (assuming an average chord length of $4R/3$).
So an arrival flux of N per year into the OC translates into
$2\times10^{4} N = 9 \times 10^{5}$ macroscopic particles from \alphacen{short} currently within the OC. Although possibly abundant within the notional bounds of our Solar System, this number decreases significantly if we consider that our practical observation limit is only $\leq10~$au from the Sun. This region is only $10^{-12}$ the volume of the OC and thus only expected to contain an object from \alphacen{short} with probability $\sim10^{-6}$.

Telescopically observable \alphacen{short} interstellars are thus expected to be rare. But what about smaller particles, which are likely more abundant and could be observed by meteor monitoring systems? Determining the rate of ejection of particles at these sizes is even more difficult. We note here that \cite{Boe2019} SFD extends to $\sim120$~m: ideally, we would extend this to millimeter-sized particles, but the uncertainty in this process is too large to accurately make any sort of prediction on this basis.

Instead, for the sake of argument, let us consider the observability of an equivalent mass rate of ejection of 100 $\mu$m particles, about the smallest meteor sizes routinely detected in Earth's atmosphere. This material could come from cometary material released from new \alphacen{short} OC comets and then ejected by planetary perturbations much as the larger comets we considered above are, or could result from the breakup of ejected macroscopic comets in space. The question of mass ejection rates at these sizes surely deserves further study but is beyond the scope of this work.

Using our smallest comet size in this range ($100$~m) and an average comet density of $400~{\rm kg~m}^{-3}$ \citep{ahearn_2011_CometDensity-400kgm3,sierks2015_CometDensity-400kgm3}, an influx of 45 \alphacen{short} bodies $100$~m in diameter translates to
$\sim 45 \times \frac{4\pi}{3}\times50^3~{\rm m}^3\times 400~{\rm kg~m}^{-3}\approx 9.4\times 10^{9}~{\rm kg}$ of \alphacen{short} material per year.
This same mass in 100 $\mu$m diameter particles would provide 
$10^{18}$ times more particles (or $4.4\times10^{19}$). The relative cross-section of the Earth to that of the Oort cloud is $\left( \frac{6378}{10^5 \times 1.5\times 10^8} {\rm ~km}\right)^2 = 1.8 \times 10^{-19}$ and so we might expect $(4.4\times 10^{19}) \times (1.8\times10^{-19}) \approx 8$
meteors from \alphacen{short} to enter the Earth's atmosphere per year as an upper limit. This number is very uncertain by any standard, but reveals that meteors from \alphacen{short} may be currently detectable at Earth in small numbers.

During the peak intensity in $\sim28,000~$yr, we expect these values to increase by a factor of $\sim10$:
an influx of $\sim 10^{3}$ macroscopic particles per year; $10^{7}$ within the OC; a $10^{-5}$ probability of one being within 10~au of the Sun; and $10^{2}$ Earth intersecting 100~$\mu$m particles per year.

To place this flux in context, we can compare it to the observed number of Solar System meteors that enter our atmosphere at these sizes. The Canadian Meteor Orbit Radar (CMOR) \citep{webster2004CMOR} is an all-sky meteor patrol radar that observes meteors above southwestern Ontario in Canada 24 hours a day down to a limiting mass of $10^{-8}$ kg, corresponding to 200 $\mu$m (diameter) sizes for a density of 1000 kg m$^{-3}$ ([574 $\mu$m, 135 $\mu$m] for densities of [100 kg m$^{-3}$, 7800 kg m$^{-3}$] corresponding to
cometary material and iron respectively). CMOR observations provide the best measurements of meteor fluxes at sizes comparable to those we are considering here. \cite{Froncisz_2020} provide the number of meteoroid orbits measured by CMOR and an integrated time-area product over 7.5 years of operation. From this, we can estimate that roughly $7\times10^{12}$ meteoroids of all types (but essentially all from our Solar System) enter Earth's atmosphere per year. As a result, only about 1 in $10^{12}$ 100~$\mu$m-sized meteors observed at Earth might be from \alphacen{short}.

Meteors from \alphacen{full} are extremely rare events, vastly outnumbered by those originating in our Solar System. Nevertheless, understanding the properties of particles that could be arriving from \alphacen{short} will aid in the detection of these elusive but potentially highly informative visitors.

\section{Conclusions}\label{sec:Conclusion}

This work examines the possibility of material from our nearest stellar neighbour \alphacen{full} arriving at our Solar System. In particular we explored the delivery of sub-mm through km-sized bodies ejected from that system by gravitational scattering within the last 100 Myr.

Our study lead us to the following conclusions:
\begin{itemize}
    \item Material from \alphacen{full} can reach and likely is already within our Solar System
    \item Most material arriving from \alphacen{full} has traveled for $<10~$Myr
    \item Material that reaches us typically left \alphacen{full} with low ($v_{\infty}<2~{\rm km~s}^{-1}$) asymptotic speeds
    \item The delivery of particles from \alphacen{full} is concentrated during a $\sim10~$Myr period, with peak intensity centered after \alphacen{short}'s closest approach ($t\approx28,000~$yr)
    \item The median velocity of the ejecta relative to the Sun at the time of close approach is $\Delta v=32.50$ km s$^{-1}$, similar to the current relative velocity of \alphacen{full} ($\Delta v=32.37$ km s$^{-1}$ \citep{SIMBAD}) 
    \item If any of this material enters the inner Solar System, its fall into the Sun's gravitational well will accelerate it to a typical heliocentric velocity $v_{hel}=53~{\rm km~s}^{-1}$ at $1~$au
    \item The expected radiant of \alphacen{full} meteors at the current time (($\alpha$, $\delta$)=($292^\circ\pm 1^\circ$, $-43^\circ \pm 2^\circ$)) largely corresponds to \alphacen{short}'s effective radiant, set by that system's velocity relative to the Sun.
    However, as \alphacen{short}'s closest approach nears, the radiant will move and have a larger spread (($\alpha$, $\delta$)=($249^\circ \pm 17^\circ$, $-61^\circ \pm 8^\circ$)) resulting from the increased range of allowable ejection speeds and directions near closest approach when the Solar System's apparent cross-section is largest.
    \item We expect that particles larger than a few microns in size are able to survive the journey from \alphacen{full}
    \item If \alphacen{short} ejects comets at a rate comparable to the current Solar System rate, we expect $\sim10^{6}$ macroscopic \alphacen{full} particles to be currently within our Solar System, though the chance of one being detectable (that is, within 10 au of our star) is only one in a million.
    \item Estimates of \alphacen{full} meteor fluxes at the Earth are extremely uncertain, but a first approximation predicts perhaps as many as $\sim 10$ detectable meteors per year in Earth's atmosphere currently, and that the current rate should increase by a factor of 10 in the next 28,000 years
\end{itemize}


A thorough understanding of the mechanisms by which material could be transferred from \alphacen{full} to the Solar System not only deepens our knowledge of interstellar transport but also opens new pathways for exploring the interconnectedness of stellar systems and the potential for material exchange across the Galaxy.


\begin{acknowledgments}
This work was supported in part by the
NASA Meteoroid Environment Office under Cooperative Agreement No. 80NSSC24M0060 and by the Natural Sciences and Engineering Research Council of Canada (NSERC) Discovery Grant program (grant No. RGPIN-2024-05200).

This research has made use of the SIMBAD database, operated at CDS, Strasbourg, France.

This work has made use of data from the European Space Agency (ESA) mission
{\it Gaia} (\url{https://www.cosmos.esa.int/gaia}), processed by the {\it Gaia}
Data Processing and Analysis Consortium (DPAC,
\url{https://www.cosmos.esa.int/web/gaia/dpac/consortium}). Funding for the DPAC
has been provided by national institutions, in particular the institutions
participating in the {\it Gaia} Multilateral Agreement.
\end{acknowledgments}


\bibliography{Gregg_IntMet}{}
\bibliographystyle{aasjournal}

\end{document}